\documentclass[aps, superscriptaddress,nofootinbib, reprint]{revtex4-2}

\usepackage{xcolor}
\usepackage[%
colorlinks=true,
linkcolor=blue,
citecolor=blue,
]{hyperref}

\usepackage[marginal]{footmisc}
\usepackage{graphicx}
\usepackage{dcolumn}
\usepackage{bm}

\usepackage{amsmath}
\allowdisplaybreaks[4] 

\usepackage[T1]{fontenc} 
\usepackage{tensor}
\usepackage{braket}  
\usepackage{bm}      
\usepackage{upgreek}
\usepackage{extarrows}

\newcommand\MaE{\mspace{2mu}\mathrm{e}\mspace{2mu}}

\newcommand\MaPI{\mspace{2mu}\uppi\mspace{2mu}}

\newcommand\MaI{\mathrm{i}}

\begin{document}

\title{Strong backreaction of gauge quanta produced during inflation}

\author{Jian-Feng He}
\email{hejianfeng@itp.ac.cn}
\affiliation{Institute of Theoretical Physics, Chinese Academy of Sciences (CAS), Beijing 100190, China}
\affiliation{School of Physical Sciences, University of Chinese Academy of Sciences, No.19A Yuquan Road, Beijing 100049, China}

\author{Kai-Ge Zhang}
\email{zhangkaige21@mails.ucas.ac.cn}
\affiliation{International Centre for Theoretical Physics Asia-Pacific, University of Chinese Academy of Sciences, 100190 Beijing, China}
\affiliation{Taiji Laboratory for Gravitational Wave Universe, University of Chinese Academy of Sciences, 100049 Beijing, China}

\author{Chengjie Fu}
\email{fucj@ahnu.edu.cn}
\affiliation{Department of Physics, Anhui Normal University, Wuhu, Anhui 241002, China}

\author{Zong-Kuan Guo}
\email{guozk@itp.ac.cn}
\affiliation{Institute of Theoretical Physics, Chinese Academy of Sciences (CAS), Beijing 100190, China}
\affiliation{School of Physical Sciences, University of Chinese Academy of Sciences, No.19A Yuquan Road, Beijing 100049, China}
\affiliation{School of Fundamental Physics and Mathematical Sciences, Hangzhou Institute for Advanced Study, University of Chinese Academy of Sciences, Hangzhou 310024, China}


\begin{abstract}
During inflation an axion field coupled to a gauge field through a Chern-Simons term can trigger the production of gauge quanta due to a tachyonic instability.
The amplification of the gauge field modes
exponentially depends on the velocity of the axion field,
which in turn slows down the rolling of the axion field when backreaction is taken into account.
To illustrate how the strength of the Chern-Simons coupling and the slope of the axion potential influence the particle production,
in this paper we consider a toy model in which the axion field is a spectator with a linear potential.
In the strong backreaction regime,
the energy density of the gauge field quasiperiodically oscillates.
The steep slope of the axion potential linearly increases the peak amplitude of the energy density
while the strong coupling linearly decreases the peak amplitude.
Additionally, we calculate the energy spectrum of gravitational waves.
\end{abstract}
\maketitle


\section{Introduction}

Inflation~\cite{Guth:1980zm, Sato:1980yn, Linde:1981mu, Albrecht:1982wi, Starobinsky:1980te} is a successful paradigm for addressing the horizon and flatness problems in the standard hot Big Bang model.
It provides an exponentially expanding phase that precedes the standard Big Bang. During the exponential expansion, regions that were initially causally connected are stretched beyond the Hubble horizon and thus become causally disconnected.
At the same time, the expansion exponentially suppresses spatial curvature. Such an expansion is typically driven by a scalar field known as the inflaton, which has large and flat potential.
Observations indicate that inflation must persist for a sufficient duration, typically about 60 {\it{e}}-folds.
During inflation, quantum fluctuations on sub-Hubble scales are stretched beyond the Hubble horizon and become classical on large scales, inducing energy density fluctuations~\cite{Starobinsky:1979ty, Mukhanov:1981xt, Hawking:1982cz, Guth:1982ec, Starobinsky:1982ee, Abbott:1984fp}.
Scalar perturbations seed large-scale structure formation and leave observable imprints in the cosmic microwave background (CMB).
Current CMB observations indicate that the power spectrum of the primordial scalar perturbations on CMB scales is nearly scale-invariant, with an amplitude $A_{\mathrm{s}}$ of approximately $2.1 \times 10^{-9}$~\cite{Planck:2018jri, Planck:2019kim, BICEP:2021xfz}.
Tensor perturbations are described as primordial gravitational waves (GWs),
which are closely tied to the energy scale of inflation.
Consequently, measuring tensor perturbations can help determine that scale.
Because tensor perturbations give rise to the B-mode polarization of the CMB,
searching for B-mode signals on large scales offers a promising way to detect primordial GWs~\cite{Alvarez:2019rhd, Shandera:2019ufi, Campeti:2019ylm, Abazajian:2019eic, Komatsu:2022nvu, Campeti:2022acx, Fujita:2022qlk, LiteBIRD:2023zmo}.
Although primordial GWs have not yet been detected, CMB observations constrain the tensor-to-scalar ratio to $r < 0.036$ at 95\,\% confidence level on the pivot scale of $0.05~\rm{Mpc}^{-1}$ \cite{BICEP:2021xfz}.
On smaller scales, these GWs can be probed through searches for a stochastic gravitational wave background using pulsar timing arrays or laser interferometers.

Given that the energy scale of inflation far exceeds that of the standard model of particle physics,
it is natural to anticipate the presence of many beyond-standard-model fields during inflation.
During inflation, scenarios in which an axion-like field acts either as the inflaton or as a spectator have been widely explored~\cite{Barnaby:2010vf, Sorbo:2011rz, Barnaby:2011vw, Barnaby:2011qe, Linde:2012bt, Mukohyama:2014gba, Namba:2015gja, Agrawal:2017awz, Ozsoy:2017blg, Agrawal:2018mrg, Papageorgiou:2019ecb, Dimastrogiovanni:2023juq, Ozsoy:2024apn, He:2024bno}.
If an axion-like field $\chi$ is coupled to a gauge field through a Chern-Simons term $\chi \tilde{F}_{\mu\nu} F^{\mu\nu}$,
it can introduce an additional mass term into the equation of motion (EoM) for the gauge field~\cite{Barnaby:2010vf, Barnaby:2011vw, Barnaby:2011qe, Sorbo:2011rz}. The mass term is generally proportional to the velocity of the axion field,
the coupling constant in the Chern-Simons term, and the inverse of the Hubble parameter.
Depending on the helicity states of the gauge field, the mass term can be either positive or negative.
Consequently, one of the helicity states may undergo a tachyonic instability, leading to exponential growth.
This significant amplification produces an additional contribution of scalar and tensor fluctuations in addition to those generated by quantum vacuum fluctuations.
The current CMB observations constrain both the amplitude and non-Gaussianity of scalar perturbations on the CMB scale and therefore constrain the gauge field amplitude during the early stages of inflation~\cite{Alam:2024fid}. However, if the axion potential includes step-like features, such as steep cliffs and gentle plateaus, only those gauge field modes that exit the horizon during the fast roll of the axion field will be significantly excited, thereby sourcing large scalar and tensor perturbations on small scales and not violating the large scale constraint.
As a result, this mechanism can produce a significant amount of primordial black holes~\cite{Garcia-Bellido:2016dkw, Garcia-Bellido:2017aan, Ozsoy:2020kat, Talebian:2022cwk, Ozsoy:2023ryl, LISACosmologyWorkingGroup:2023njw} and the observable GWs on small scales~\cite{Giare:2020vhn, Campeti:2020xwn, LISACosmologyWorkingGroup:2022jok, Corba:2024tfz, Garcia-Bellido:2023ser}.
Furthermore, some studies suggest that such GWs could potentially explain the recent results from pulsar timing arrays~\cite{NANOGrav:2023hvm, Figueroa:2023zhu, Unal:2023srk, Niu:2023bsr, Maiti:2024nhv}.

Note that the gauge quanta produced through the tachyonic instability can decelerate the axion field, thereby suppressing further the particle production. This process is commonly referred to as \emph{backreaction} in the literature.
Typically, there is a delay between the change in the velocity of the axion field and the backreaction term~\cite{Domcke:2020zez, Peloso:2022ovc}.
If backreaction is sufficiently strong,
it can induce oscillations in the velocity of the axion field.
Additionally, much of the literature numerically explores the effects of backreaction~\cite{Cheng:2015oqa, Notari:2016npn, DallAgata:2019yrr, Gorbar:2021rlt, Durrer:2023rhc, vonEckardstein:2023gwk, Iarygina:2023mtj, Garcia-Bellido:2023ser, Caravano:2022epk, Galanti:2024jhw, Caravano:2024xsb, Figueroa:2024rkr, Sharma:2024nfu}.
However, previous studies have primarily focused on axion potential with varying slopes, which, while more realistic, complicates the analysis of how the slope affects the particle production.
Furthermore, the impact of the coupling strength has not been explored in detail.
Additionally, it remains unclear whether there is an upper bound on the particle production in the strong backreaction regime.
In this work, we examine a toy model in which the axion acts as a spectator with a linear potential.
This simplified treatment allows us to better understand how the slope of the potential influences the particle production. We use the norm of the gauge field to quantify the particle production and investigate its upper limit.
Since backreaction constrains the maximum particle production, it also limits the production of sourced GWs. We discuss these constraints in detail in the context of our model.

This paper is organized as follows.
In Section~\ref{sec: model}, we describe the model employed in this work, which consists of an inflaton with a Starobinsky potential, a spectator axion-like field with a linear potential, and a gauge field coupled to the axion field via a Chern-Simons term $\chi \tilde{F}_{\mu\nu} F^{\mu\nu}$.
In Section~\ref{sec: enhance_gauge_particle}, we analyze the influences of the coupling constant and the slope of the axion potential on the particle production, in both the weak and strong backreaction regimes.
In Section~\ref{sec: sourced_gws}, we calculate the energy spectrum of GWs generating from the gauge field in the strong backreaction regime.
In Section~\ref{sec: conclusion}, we present our conclusions. Throughout this paper, we use unit $\hbar = c = 1$, and use $M_{\mathrm{pl}} \equiv (8 \MaPI G)^{-1/2}$ to represent the reduced Planck mass.

\section{Model}
\label{sec: model}

We consider an inflationary model consisting of an inflaton, a spectator axion, and a gauge field, where the axion field is coupled to the gauge field through a Chern-Simons term. The Lagrangian for our model is given by
\begin{eqnarray}
  \mathcal{L}
  &=& \frac{M_{\mathrm{pl}}^{2}}{2} R
  - \frac{1}{2} \partial_{\mu} \phi \partial^{\mu} \phi - V(\phi)
  - \frac{1}{2} \partial_{\mu} \chi \partial^{\mu} \chi - U(\chi) \nonumber \\
  && - \frac{1}{4} F^{\mu\nu} F_{\mu\nu}
  - \frac{1}{4 f} \chi \tilde{F}^{\mu\nu} F_{\mu\nu},
\label{eq: L_axion_RFF}
\end{eqnarray}
where $\phi$ is the inflaton, $\chi$ is the axion field, $F_{\mu\nu} \equiv \partial_{\mu} A_{\nu} - \partial_{\nu} A_{\mu}$ and $\tilde{F}^{\mu\nu} \equiv \frac{\eta^{\mu\nu\alpha\beta} F_{\alpha\beta}}{2 \sqrt{-g}}$ is its dual, with the totally antisymmetric tensor $\eta^{\mu\nu\alpha\beta}$ defined such that $\eta^{0123} = 1$. Here $f$ is the decay constant with a mass dimension of one, and it should satisfy $H \ll f \ll M_{\mathrm{pl}}$ \cite{Barnaby:2011pe}. Throughout this paper, we use the spatially flat Friedmann-Lematre-Robertson-Walker (FLRW) metric, $\mathrm{d}s^{2} = -\mathrm{d}t^{2} + a^{2}(t)\mathrm{d} \bm{x}^{2} = a^{2}(\tau)(-\mathrm{d}\tau^{2} + \mathrm{d} \bm{x}^{2})$, where $\tau$ denotes the conformal time. We assume the inflaton slowly rolls, and its potential is consistent with the current observation constraints. For example, the Starobinsky potential $V(\phi) = V_{0} \left[ 1 - \exp\left( - \sqrt{2 / 3} \phi \right) \right ]^{2}$, with $V_{0} = 9.75 \times 10^{-11} M_{\mathrm{pl}}^{4}$, satisfies all the necessary requirements.

We assume the axion field acts as a spectator, meaning its potential is significantly smaller than that of the inflaton. Consequently, its evolution does not influence the scale factor $a$. In realistic models, the potential of the axion $\chi$ is generally nonlinear and includes features such as cliffs to generate large perturbations at small scales. Due to the varying slope of the potential, it is challenging to independently investigate the influence of the slope. However, the potential can always be expanded using a Taylor series, and the lowest-order term that affects the evolution of the axion field is the linear term. Therefore, it is reasonable to approximate the fast-roll stage with a linear potential
\begin{equation}
  U(\chi) = c \chi,
\end{equation}
where $c$ is a constant and has a mass dimension of 3. This toy model enables us to focus on the influence of the slope while avoiding additional complexities.

\begin{figure}[tbp]
  \includegraphics[width=.48\textwidth]{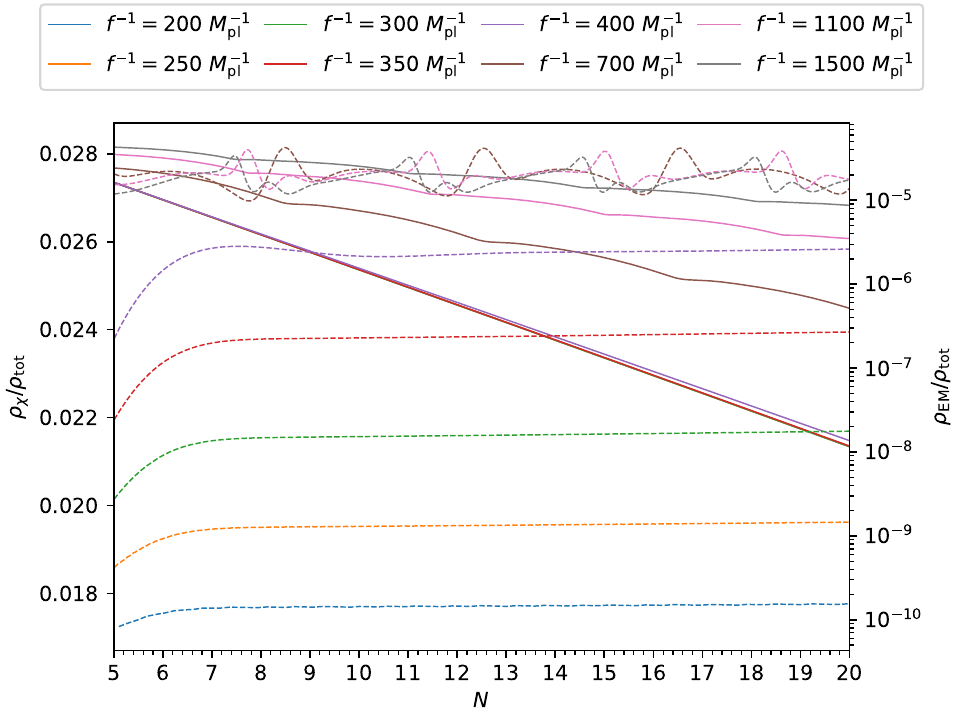}
  \caption{Evolution of $\rho_{\chi} / \rho_{\mathrm{tot}}$ and $\rho_{\mathrm{EM}} / \rho_{\mathrm{tot}}$ for different coupling constants with $c = 2 \times 10^{-12} \, M_{\mathrm{pl}}^{3}$. The solid lines represent $\rho_{\chi} / \rho_{\mathrm{tot}}$ and the dashed lines represent $\rho_{\mathrm{EM}} / \rho_{\mathrm{tot}}$.}
  \label{fig: rhochi_rho_EM_varalp}
\end{figure}

\begin{figure}[tbp]
  \includegraphics[width=.48\textwidth]{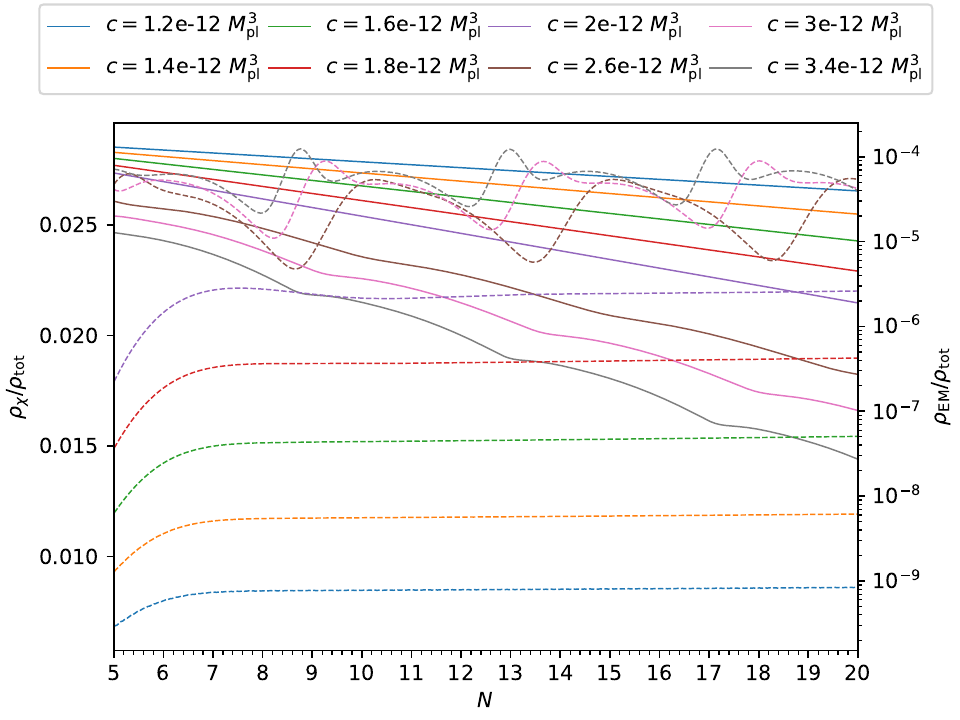}
  \caption{Evolution of $\rho_{\chi} / \rho_{\mathrm{tot}}$ and $\rho_{\mathrm{EM}} / \rho_{\mathrm{tot}}$ for different slopes with $f^{-1} = 400 \, M_{\mathrm{pl}}$. The solid lines represent $\rho_{\chi} / \rho_{\mathrm{tot}}$ and the dashed lines represent $\rho_{\mathrm{EM}} / \rho_{\mathrm{tot}}$.}
  \label{fig: rhochi_rho_EM_varcslope}
\end{figure}

The Lagrangian~\eqref{eq: L_axion_RFF} combined with the spatially flat FLRW metric yields the following EoM of the background quantities
\begin{align}
  \label{eq: a_eom_1}
  & H^{2} = \frac{1}{3 M_{\mathrm{pl}}^{2}} \rho\,, \\
  \label{eq: a_eom_2}
  & \frac{\ddot{a}}{a} + \frac{1}{2} \left( \frac{\dot{a}}{a} \right)^{2}
  = - \frac{1}{2 M_{\mathrm{pl}}^{2}} P\,, \\
  \label{eq: phi_eom}
  & \ddot{\phi} + 3 H \dot{\phi} + V_{,\phi} = 0\,, \\
  \label{eq: chi_eom}
  & \ddot{\chi} + 3 H \dot{\chi} + U_{,\chi} = \frac{1}{f}
  \braket{\bm{E} \cdot \bm{B}},
\end{align}
where the overdots denote the derivative with respect to the cosmic time $t$. $\bm{E}$ and $\bm{B}$ represent the electric and magnetic fields associated with the gauge field, defined by \cite{Barnaby:2010vf, Sorbo:2011rz, Garcia-Bellido:2023ser}
\begin{align}
  & E_{i}(t) = - \dot{A}_{i} / a\,, \\
  & B_{i}(t) = \varepsilon_{ijk} \partial_{j} A_{k} / a^{2}\,.
\end{align}
The angle bracket $\braket{\cdots}$ denotes the ensemble average of the fields, which can be computed by evaluating the vacuum average of the corresponding field operators.
The averaged term encodes backreaction of the quantum field on the classic background. The total energy density $\rho$ in Eq.~\eqref{eq: a_eom_1} and the total pressure $P$ in Eq.~\eqref{eq: a_eom_2} are given by
\begin{align}
  \label{eq: rho_tot}
  \rho =&
  \frac{1}{2} \dot{\phi}^{2}
  + \frac{1}{2} \dot{\chi}^{2}
  + V(\phi) + U(\chi)
  + \frac{1}{2} \braket{ \bm{E}^{2} + \bm{B}^{2} }, \\
  \label{eq: p_tot}
  P =&
  \frac{1}{2} \dot{\phi}^{2} +  \frac{1}{2} \dot{\chi}^{2}
  - V(\phi) - U(\chi)
  + \frac{1}{6} \braket{\bm{E}^{2} + \bm{B}^{2}}.
\end{align}
Figures \ref{fig: rhochi_rho_EM_varalp} and \ref{fig: rhochi_rho_EM_varcslope} display the evolution of the axion energy density $\rho_{\chi} = \frac{1}{2} \dot{\chi}^{2} + U(\chi)$ and the gauge field energy density $\rho_{\mathrm{EM}} = \frac{1}{2} \braket{E^{2} + B^{2}}$, across the parameter space of interest. All energy densities are normalized by the total energy density. In all the cases, the axion energy density remains small compared to the inflaton, confirming that the axion behaves as a spectator field. Moreover, the gauge field energy density is even smaller than that of the axion, indicating that its backreaction on the evolution of the scale factor is negligible. Notably, Fig. \ref{fig: rhochi_rho_EM_varalp} illustrates that strong backreaction effects can significantly slow down the axion.

We decompose the gauge field operator $\hat{A}_{i}$ in the Fourier space as
\begin{align}
  & \hat{A}_{i}(t, \bm{x})
  =  \int \frac{\mathrm{d}^{3} \bm{k}}{(2 \MaPI)^{3 / 2}}
  \MaE^{ \MaI \bm{k} \cdot \bm{x} } \hat{A}_{i}(t, \bm{k}) \\
  \label{eq: Ak_decompose}
  & =  \int \frac{\mathrm{d}^{3} \bm{k}}{(2 \MaPI)^{3 / 2}}
  \sum_{\lambda=\pm} \MaE^{ \MaI \bm{k} \cdot \bm{x} }
    \epsilon^{\lambda}_{i}(\hat{k})
    A^{\lambda}(t, \bm{k}) \hat{a}^{\lambda}(\bm{k}) + \text{h.c.},
\end{align}
where $\lambda = +, -$ denotes the polarization index, $\hat{k} \equiv \bm{k}/k$ is a unit vector with the same direction as the $\bm{k}$, $k \equiv |\bm{k}|$ is the norm, $\hat{a}(\bm{k})$ and $\hat{a}^{\dag}(\bm{k})$ are annihilation and creation operators satisfying $[\hat{a}^{\lambda}(\bm{k}), \hat{a}^{\lambda'\dag}(\bm{k'})] = \delta^{\lambda\lambda'} \delta^{(3)}(\bm{k} - \bm{k}')$, and $\epsilon_{i}^{\lambda}(\hat{k})$ is the polarization vector basis satisfying \cite{Sorbo:2011rz}
\begin{align}
  & k_{i} \epsilon^{\pm}_{i}(\hat{k}) = 0,
  ~ \varepsilon_{ijk} k_{j} \epsilon_{k}^{\pm}(\hat{k})
  = \mp \MaI k \epsilon^{\pm}_{i}(\hat{k}), \\
  & \epsilon_{i}^{\pm}(\hat{k}) \epsilon_{i}^{\pm}(\hat{k}) = 0,
  ~ \epsilon_{i}^{\pm}(\hat{k}) \epsilon_{i}^{\mp}(\hat{k}) = 1.
\end{align}
Then, by the definition of $\bm{E}$ and $\bm{B}$, one can compute the ensemble averages in Eqs.~\eqref{eq: chi_eom}, \eqref{eq: rho_tot} and \eqref{eq: p_tot} as
\begin{align}
  \label{eq: ensemble_EB}
  & \braket{\bm{E} \cdot \bm{B}} =
  - \frac{1}{4 \MaPI^{2} a^{4}}
  \sum_{\lambda=\pm} \lambda \int_{0}^{\infty} \mathrm{d}k k^{3}
  \frac{\mathrm{d} }{\mathrm{d} \tau} | A^{\lambda}(k) |^{2}, \\
  \label{eq: ensemble_EE}
  & \braket{E^{2}} = \frac{1}{2 \MaPI^{2} a^{4}}
  \sum_{\lambda=\pm} \int_{0}^{\infty}\mathrm{d}k
  k^{2} \left| \frac{\mathrm{d} A^{\lambda}(k)}{\mathrm{d} \tau} \right|^{2}, \\
  \label{eq: ensemble_BB}
  & \braket{B^{2}} = \frac{1}{2 \MaPI^{2} a^{4}}
  \sum_{\lambda=\pm} \int_{0}^{\infty} \mathrm{d}k
  k^{4} |A^{\lambda}(k)|^{2}.
\end{align}

\section{Strong backreaction}
\label{sec: enhance_gauge_particle}

The gauge field mode functions $A^{\lambda}(\bm{k})$ in Eq.~\eqref{eq: Ak_decompose} obey the following EoM
\begin{equation}
  \label{eq: Ak_eom_t}
  \ddot{A}^{\pm}(\bm{k}) + H \dot{A}^{\pm}(\bm{k})
  + \left( \frac{k^{2}}{a^{2}} \mp \frac{k}{a} \frac{1}{f} \dot{\chi} \right)
  A^{\pm}(\bm{k}) = 0.
\end{equation}
When the coefficient of the $A^{\pm}(\bm{k})$ term is negative, the corresponding mode with $\lambda = \mathrm{sign}(\dot{\chi})$ undergoes exponential growth, a phenomenon known as tachyonic instability. Without loss of generality, we focus on the case $\mathrm{sign}(\dot{\chi}) = -$ in this work.
The EoM~\eqref{eq: Ak_eom_t} can be rewritten in terms of the conformal time as~\cite{Sorbo:2011rz, Garcia-Bellido:2023ser}
\begin{equation}
  \label{eq: Ak_eom}
  A^{\pm''}(\bm{k}) + (k^{2} \pm 2 \xi aH k) A^{\pm}(\bm{k}) = 0.
\end{equation}
Here, the parameter $\xi=-\dot{\chi}/(2 f H)$ is a dimensionless parameter that encodes information about the inflationary energy scale and the velocity of the axion field. The initial condition of the mode functions is determined by the Bunch-Davies vacuum
\begin{equation}
  \label{eq: bd_vacuum}
  A^{\pm}_{\mathrm{BD}}(\bm{k}) \equiv \left. A(\tau, \bm{k}) \right|_{k \gg aH}
  = \frac{1}{\sqrt{2 k}} \MaE^{ - \MaI k \tau }.
\end{equation}
As the Universe expands, when ${k}/({aH}) < 2 \xi$, the mode $k$ with polarization $\lambda = -$ will undergo exponential growth. Once the mode $k$ is well outside the horizon (i.e., $k \ll aH$), the coefficient of the $A^{\pm}(\bm{k})$ term in Eq.~\eqref{eq: Ak_eom_t} vanishes, and the mode freezes. Therefore, the final value of the norm $\left| A^{\pm}(\bm{k}) \right|^{2}$ is mainly determined by the parameter $\xi$ during the interval from ${k}/({aH}) = 2 \xi$ until $k \lesssim aH$. If the velocity of the axion field remains nearly constant, the solution is given by the irregular Coulomb function~\cite{Sorbo:2011rz, Garcia-Bellido:2023ser}.
When the axion's velocity varies only slowly, the Wentzel–Kramers–Brillouin (WKB) approximation can be applied~\cite{Ozsoy:2020kat}.
However, if the velocity of the axion changes rapidly, both of these methods fail, necessitating a full numerical computation.

The EoM for the gauge field, Eq.~\eqref{eq: Ak_eom}, demonstrates that the enhancement of the gauge field due to the tachyonic instability is determined by the velocity of the axion field. Conversely, the EoM for the axion field, Eq.~\eqref{eq: chi_eom}, shows the backreaction term $f^{-1} \braket{\bm{E} \cdot \bm{B}}$ can influence the dynamics of the axion. If this backreaction term becomes comparable to the slope term $U_{,\chi}$, the axion will decelerate, leading to a reduction in the particle production. As a result, the backreaction term $f^{-1} \braket{\bm{E} \cdot \bm{B}}$ decreases, allowing the axion to accelerate once more. Specifically, the deceleration condition is expressed as $\mathrm{d} |\dot{\chi}| / \mathrm{d}t < 0$, which, for $\dot{\chi} < 0$, is equivalent to $f^{-1} \braket{\bm{E} \cdot \bm{B}} > 3 H \dot{\chi} + U_{,\chi}$. Typically, the combination $3H \dot{\chi} + U_{,\chi}$ is of the same order of magnitude as $U_{,\chi}$. Therefore, we arrive at the following approximation for the strong backreaction condition
\begin{equation}
  \label{eq: axion_decce_condition}
  f^{-1} \braket{\bm{E} \cdot \bm{B}} \sim U_{,\chi}.
\end{equation}

\begin{figure}[tbp]
  \includegraphics[width=.48\textwidth]{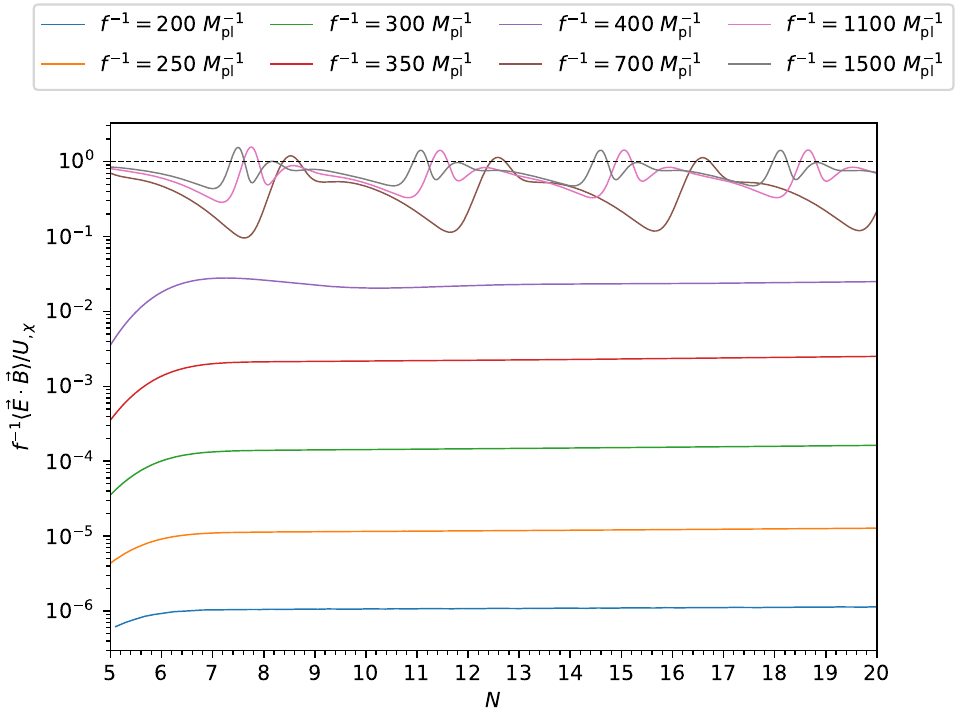}
  \caption{Evolution of the ratio between the backreaction term and the axion slope $f^{-1}\braket{\bm{E} \cdot \bm{B}} / U_{,\chi}$ for different coupling constants with $c = 2 \times 10^{-12} \, M_{\mathrm{pl}}^{3}$. The black dashed indicates the threshold where the bacreaction term equals the axion slope, i.e. $f^{-1}\braket{\bm{E} \cdot \bm{B}} = U_{,\chi}$.}
  \label{fig: finvEB_over_slope_varalp}
\end{figure}

\begin{figure}[tbp]
  \includegraphics[width=.48\textwidth]{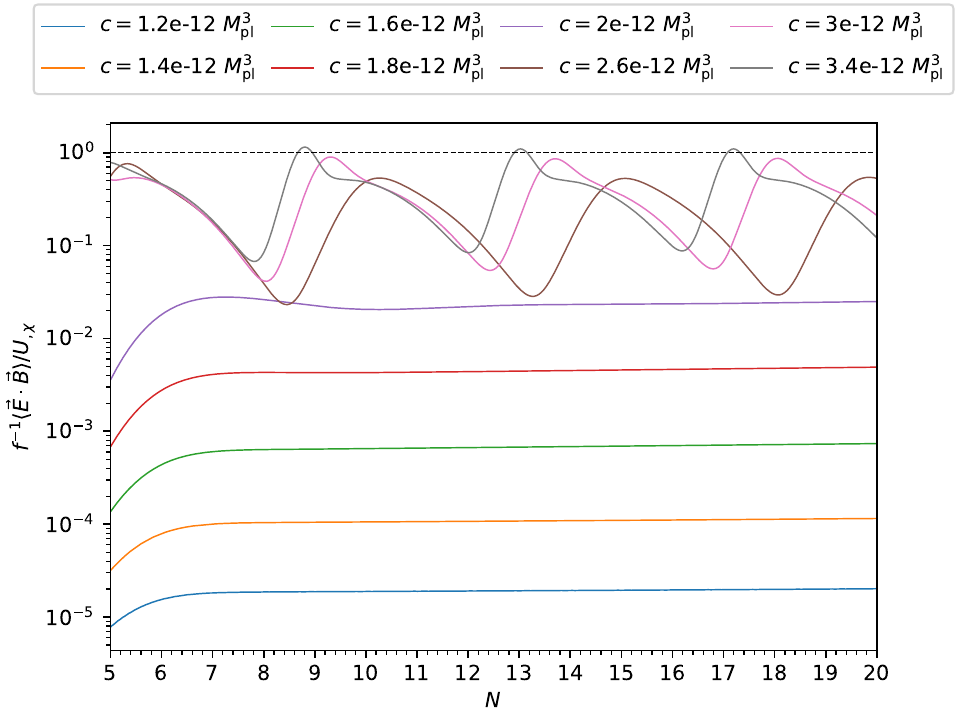}
  \caption{Evolution of the ratio between the backreaction term and the axion slope $f^{-1}\braket{\bm{E} \cdot \bm{B}} / U_{,\chi}$ for different slopes with $f^{-1} = 400 \, M_{\mathrm{pl}}$. The black dashed indicates the threshold where the bacreaction term equals the axion slope, i.e. $f^{-1}\braket{\bm{E} \cdot \bm{B}} = U_{,\chi}$.}
  \label{fig: finvEB_over_slope_varcslope}
\end{figure}

Generally, the deceleration stage induced by backreaction is brief, typically lasting less than one {\it{e}}-folding number in the strong backreaction regime. Eq.~\eqref{eq: axion_decce_condition} provides an intuitive framework for understanding how $f$ and $U_{,\chi}$ influence the maximum particle production.
Since $\braket{\bm{E} \cdot \bm{B}}$ is approximately proportional to $|A^{-}|^{2}$ (with the minus mode being the enhanced one) and Eq.~\eqref{eq: axion_decce_condition} shows that $\braket{\bm{E} \cdot \bm{B}}$ is roughly proportional to $U_{,\chi} f$, one can expect that the peak value of $|A^{-}|^{2}$ is also proportional to $U_{,\chi} f$.
Figures \ref{fig: finvEB_over_slope_varalp} and \ref{fig: finvEB_over_slope_varcslope} show the evolution of the ratio between the backreaction term and the axion slope, $f^{-1}\braket{\bm{E} \cdot \bm{B}} / U_{,\chi}$, across the relevant parameter space. Those figures demonstrate that when the backreaction term $f^{-1} \braket{\bm{E} \cdot \bm{B}}$ is much smaller than the slope $U_{\chi}$, its influence on the dynamics of the system is negligible. However, when the backreaction term and the slope are of comparable magnitude, the backreaction induces strong oscillations in particle production, a phenomenon that we will analyze in detail later in this section. Therefore, Figs. \ref{fig: finvEB_over_slope_varalp} and \ref{fig: finvEB_over_slope_varcslope} provide strong support for the validity of condition \eqref{eq: axion_decce_condition}.

To numerically solve the background equations~\eqref{eq: a_eom_2}-\eqref{eq: chi_eom} and the perturbation equation~\eqref{eq: Ak_eom_t}, we implemented a C program utilizing the Runge-Kutta algorithm. In our program, the background evolution is computed by the 4th order Runge-Kutta algorithm, while the gauge field evolution is computed by the 2nd order Runge-Kutta algorithm. We employ a normalized cosmic time variable defined as $\tilde{t} = \sqrt{V_{0}} t / M_{\mathrm{pl}}$. With the new time variable, the total duration of inflation, $\tilde{t}_{\mathrm{tot}}$, is on the order of 100. For each mode $k$, the evolution begins deep inside the horizon at $k = 100 \, aH$, with Bunch-Davies vacuum Eq.~\eqref{eq: bd_vacuum} serving as the initial condition. To accurately resolve the behavior in the deep subhorizon regime, the global time step is set to $\Delta \tilde{t} = 10^{-5}$. In all subsequent calculations, we define the end of inflation by the condition $\epsilon = 1$. To ensure that the axion remains a spectator, we set $U(\chi_{\mathrm{ini}}) = 0.02 V(\phi_{\mathrm{ini}})$, with $\phi_{\mathrm{ini}} \simeq 4.5 M_{\mathrm{pl}}$ being the field value at the time when the CMB pivot scale exits the horizon.


\subsection{Influence of the coupling constant}

In this subsection, to investigate the influence of the coupling constant, we vary $f$ while keeping the slope of the axion potential fixed. The parameter $c$, which determines the slope, is set to $c = 2 \times 10^{-12} \, M_{\mathrm{pl}}^{3}$.

\begin{figure}[tbp]
  \includegraphics[width=.48\textwidth]{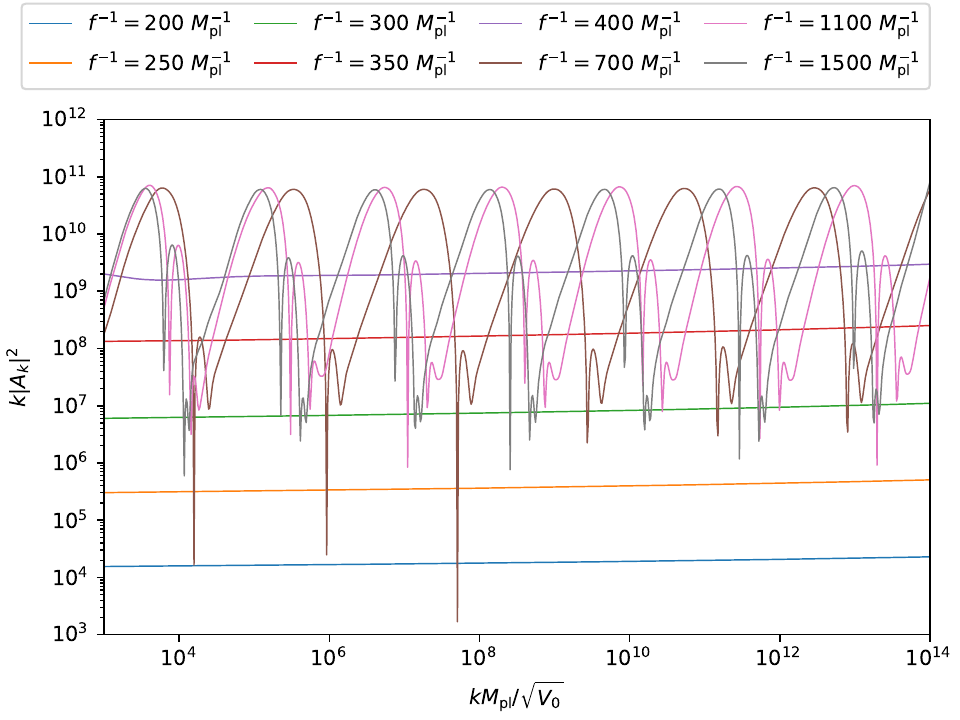}
  \caption{Gauge field spectrum at the end of inflation for different coupling constants with $c = 2 \times 10^{-12} \, M_{\mathrm{pl}}^{3}$. The first five lines appear as straight lines, indicating that in these cases backreaction is weak and does not significantly affect the particle production. When $f^{-1}$ is increased uniformly from $200.0 \, M_{\mathrm{pl}}^{-1}$ to $400.0 \, M_{\mathrm{pl}}^{-1}$, the gauge field spectrum increases uniformly on a logarithmic scale,  reflecting an exponential dependence on the coupling coefficient $f^{-1}$. Further increasing $f^{-1}$ causes backreaction to become strong enough to influence the velocity of the axion field, leading to oscillations in the gauge field spectrum. The last three curves illustrate that in the strong backreaction regime, increasing $f^{-1}$ no longer significantly enhances the maximum gauge field spectrum. Nevertheless, since particles can be produced more quickly, the oscillation frequency increases for larger $f^{-1}$.}
  \label{fig: Aqspec_varalp}
\end{figure}

\begin{figure}[tbp]
  \includegraphics[width=.48\textwidth]{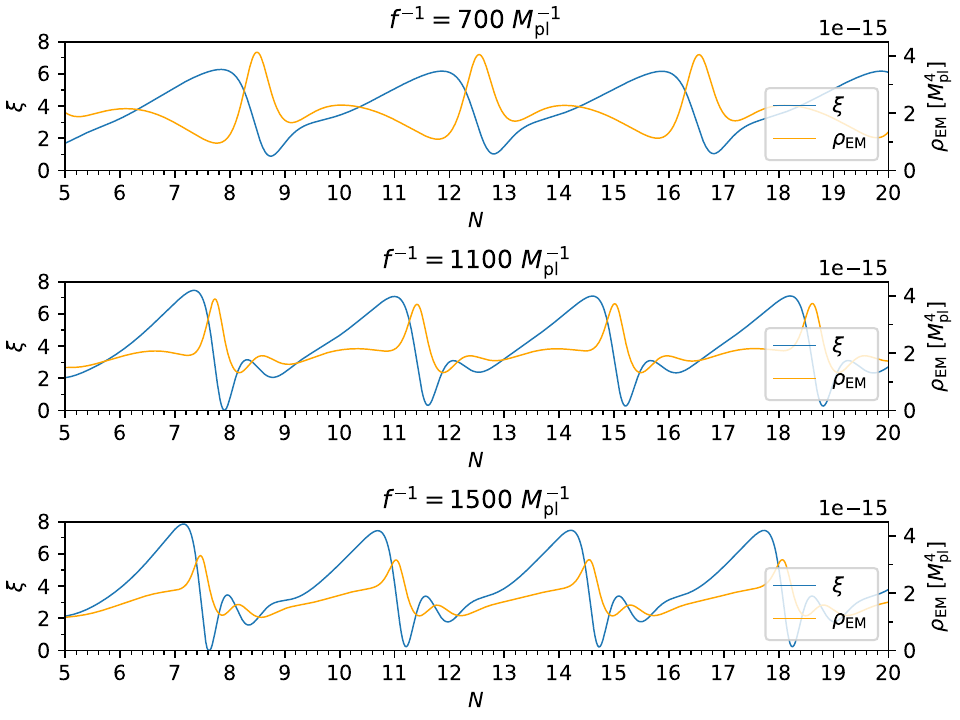}
  \caption{Evolution of $\xi$ and $\rho_{\mathrm{EM}}$ for different coupling constants in the strong backreaction regime with $c = 2 \times 10^{-12} \, M_{\mathrm{pl}}^{3}$. The blue line denotes the evolution of the parameter $\xi$ and the orange line denotes the evolution of the energy density of the gauge field. A clear time delay emerges between their respective peaks, with $\xi$ always reaching its maximum before $\rho_{\mathrm{EM}}$, indicating that the particle production is not instantaneous. As $f^{-1}$ increases, this delay shortens. Although a larger $f^{-1}$ accelerates the particle production, the accompanying faster growth of backreaction prevents a substantial enhancement of the final gauge field spectrum. Moreover, for larger $f^{-1}$, the peak of $\rho_{\mathrm{EM}}$ is approximately linearly suppressed, corroborating our earlier conclusion in Sec. \ref{sec: enhance_gauge_particle} that an increased $f^{-1}$ ultimately reduces the overall particle production.}
  \label{fig: Ne_xi_rhoEM_br_varalp}
\end{figure}

Fig.~\ref{fig: Aqspec_varalp} shows the normalized gauge field spectrum $k |A_{k}|^{2}$ at the end of inflation for various coupling constants.
We normalize by $k$ because of $1 / \sqrt{k}$ in Eq.~\eqref{eq: bd_vacuum}, which ensures all the vacuum values of $k| A_{\mathrm{BD}}^{\pm} |^{2}$ remain $1/2$ . Fig.~\ref{fig: Ne_xi_rhoEM_br_varalp} shows the evolution of the parameter $\xi$ and the energy density of the gauge field in the strong backreaction regime for different coupling constants. When the coefficient $f^{-1}$ is small,
produced particles do not significantly affect the evolution of the axion field. Consequently, the velocity of the axion remains nearly unchanged, producing a fairly flat gauge field spectrum. However, for larger values of $f^{-1}$, the particle production becomes more efficient
and backreaction correspondingly grows.
Once the backreaction term $f^{-1} \braket{\bm{E} \cdot \bm{B}}$ becomes comparable to the potential slope $U_{,\chi}$, the axion decelerates,
causing backreaction to weaken. As the backreaction subsides, the axion can accelerate again. The repetition of this cycle leads to the oscillatory behavior observed in $\xi$ and in the energy density $\rho_{\mathrm{EM}} = \frac{1}{2} \braket{E^{2} + B^{2}}$, as shown in Fig.~\ref{fig: Ne_xi_rhoEM_br_varalp}.

From Fig.~\ref{fig: Aqspec_varalp}, one can observe that when backreaction is weak, increasing the coefficient $f^{-1}$ leads to an exponential increase in the gauge quanta production. However, once the system enters the strong backreaction regime, the spectrum begins to oscillate and further increases $f^{-1}$ no longer exponentially increasing the maximum particle production. Instead, it will slightly suppress the particle production. Although this suppression is difficult to discern in Fig.~\ref{fig: Aqspec_varalp} due to the logarithmic scaling, it becomes apparent from the decreasing energy density shown in Fig.~\ref{fig: Ne_xi_rhoEM_br_varalp}.
Approximately, increasing $f^{-1}$ will linearly decrease the energy density. The insensitivity of the spectrum peak value to the coupling strength in the strong backreaction regime can be understood by examining the evolution of the parameter $\xi$ and the energy density $\rho_{\mathrm{EM}}$. In the strong backreaction regime, increasing the coefficient $f^{-1}$ leads to a larger maximum $\xi$ and a shorter delay between the peaks of $\xi$ and $\rho_{\mathrm{EM}}$. A higher peak in $\xi$ implies that, over a given time interval, more particles are produced. However, the reduced delay means that backreaction occurs earlier, leaving less time for the particle production at high efficiency.
These two competing effects tend to balance each other, preventing the peak value of the gauge field spectrum from continuing to grow exponentially.

\subsection{Influence of the potential slope}

In this subsection, to investigate the influence of the potential slope, we vary $c$ while keeping the coupling strength fixed as $f^{-1} = 400 \, M_{\mathrm{pl}}$. To ensure that the axion remains a spectator field, we impose $U(\chi_{\mathrm{ini}}) = 0.02 \, V(\phi_{\mathrm{ini}})$ for each choice of the slope.

\begin{figure}[tbp]
  \includegraphics[width=.48\textwidth]{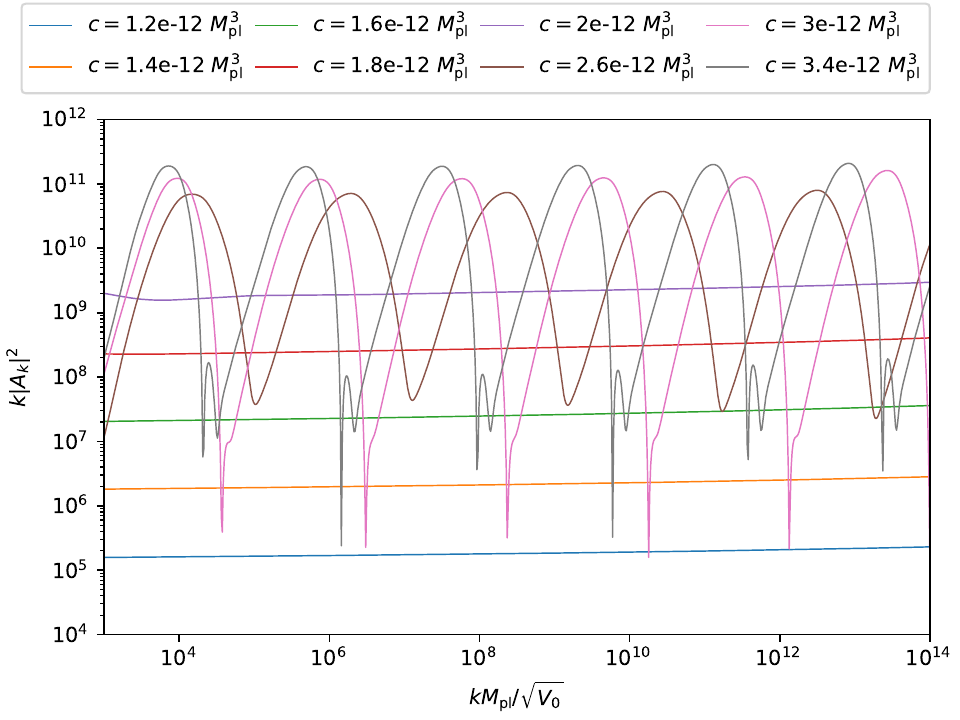}
  \caption{Gauge field spectrum at the end of inflation with different slopes, where we fix the coupling constant as $f^{-1} = 400 \, M_{\mathrm{pl}}$. In the weak backreaction region, the behavior is similar to the Fig. \ref{fig: Aqspec_varalp}, where the gauge field spectrum appears as a straight line and exhibits an exponential dependence on the slope $c$. However, in the strong backreaction regime, the behavior differs from Fig. \ref{fig: Aqspec_varalp}. Specifically, when the slope of the potential increases, both the oscillation frequency and the maximum value of the gauge field spectrum increase, although the magnitude of this enhancement is much smaller than in the weak backreaction regime.}
  \label{fig: Aqspec_varcslope}
\end{figure}

\begin{figure}[tbp]
  \includegraphics[width=.48\textwidth]{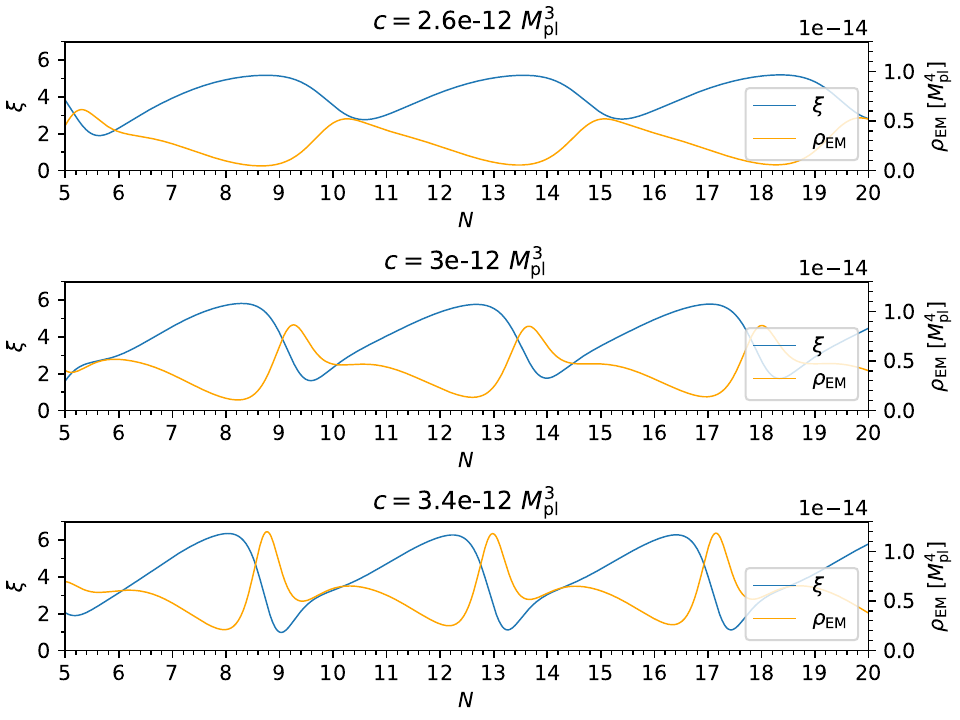}
  \caption{Evolution of $\xi$ and $\rho_{\mathrm{EM}}$ for different slopes in the strong backreaction regime, where we fix the coupling constant as $f^{-1} = 400 \, M_{\mathrm{pl}}$. The blue line denotes the evolution of the parameter $\xi$ and the orange line denotes the evolution of the energy density of the gauge field $\rho_{\mathrm{EM}}$. Similar to Fig.~\ref{fig: Ne_xi_rhoEM_br_varalp}, there is a delay between $\xi$ and $\rho_{\mathrm{EM}}$, and the delay becomes shorter for larger slopes. However, different from $f^{-1}$, as the slope increases, the peak value of the energy density will approximately linearly increase, indicating that more particles are produced. This explains the increase in the peak value of the gauge field spectrum seen in the strong backreaction regime in Fig.~\ref{fig: Aqspec_varcslope}.}
  \label{fig: Ne_xi_rhoEM_br_varcslope}
\end{figure}

Fig.~\ref{fig: Aqspec_varcslope} shows the normalized gauge field spectrum $k |A_{k}|^{2}$ at the end of inflation for different slopes of the axion potential, while Fig.~\ref{fig: Ne_xi_rhoEM_br_varcslope} displays the evolution of $\xi$ and the gauge field energy density in the strong backreaction regime. Similarly to the findings in the previous subsection, a small slope leads to weak backreaction, which does not significantly affect the velocity of the axion field, resulting in a straight line in the logarithmic plot of the gauge field spectrum. By contrast, a larger slope causes faster rolling of the axion and strengthens backreaction.
Increasing the slope further eventually makes the backreaction term comparable to the potential slope, causing both the axion velocity and the gauge field energy density to decrease. As the backreaction term then subsides, the axion field accelerates once more. Repetition of this cycle leads to oscillations in both $\xi$  and the gauge field energy density,
as seen in Fig.~\ref{fig: Ne_xi_rhoEM_br_varcslope}.

In the weak backreaction regime, the slope of the axion potential influences the gauge field spectrum in a manner similar to the coupling constant: increasing either the slope $U_{,\chi}$ or the coupling coefficient $f^{-1}$ leads to an exponential increase in the gauge field spectrum. However, the behavior changes in the strong backreaction regime. Fig.~\ref{fig: Aqspec_varcslope} demonstrates that even in the strong backreaction regime, increasing the slope can still enhance the peak value of the spectrum, although the efficiency is much lower compared to the weak backreaction regime; Fig.~\ref{fig: Ne_xi_rhoEM_br_varcslope} shows that increase the slope can approximately linearly increase the peak value of the energy density of the gauge field. Fig.~\ref{fig: Ne_xi_rhoEM_br_varcslope} also displays a pattern similar to Fig.~\ref{fig: Ne_xi_rhoEM_br_varalp}, in which increasing the slope can increase the peak value of the $\xi$ and decrease the delay between $\xi$ and $\rho_{\mathrm{EM}}$. This clarifies why, in the strong backreaction regime, increasing the slope no longer yields an exponential increase in the gauge field spectrum. On the other hand, because a larger slope $U_{,\chi}$ permits stronger backreaction (as indicated by Eq.~\eqref{eq: axion_decce_condition}), the maximum gauge field spectrum can still grow. Therefore, in realistic models, it is plausible to have a potential with a larger slope to achieve a more pronounced gauge field spectrum.


\section{Gravitational waves}
\label{sec: sourced_gws}

Tensor perturbations $h_{ij}$ of the FLRW metric are \cite{Sorbo:2011rz}
\begin{equation}
  \mathrm{d}s^{2} = a^{2}(\tau)
  \left[ - \mathrm{d}\tau^{2} + (\delta_{ij}
  + h_{ij}(\tau, \bm{x}))\mathrm{d}x_{i} \mathrm{d}x_{j} \right],
\end{equation}
where $h_{ij}$ is traceless $\sum_{i} h_{ii} = 0$ and transverse $\sum_{j}\partial_{j}h_{ij} = 0$. The EoM of the sourced $h_{ij}$ is
\begin{equation}
  h_{ij}'' + 2 \frac{a'}{a} - \Delta h_{ij}
  = \frac{2}{M_{\mathrm{pl}^{2}}}
  \tensor{{\Pi}}{_{ij}^{lm}} T_{lm}^{\mathrm{EM}},
\end{equation}
where $\tensor{{\Pi}}{_{ij}^{lm}} = \Pi^{i}_{l} \Pi^{j}_{m} - \frac{1}{2} \Pi_{ij} \Pi^{lm}$ is a traceless and transverse projector, with $\Pi_{ij} = \delta_{ij} - \partial_{i} \partial_{j} / \Delta$. $T_{lm}^{\mathrm{EM}}$ is the spatial component of the energy-momentum tensor of the gauge field and has the form $T_{ij}^{\mathrm{EM}} = -a^{2}(E_{i}E_{j} + B_{i} B_{j}) + (\cdots) \, \delta_{ij}$. Since $\tensor{{\Pi}}{_{ij}^{lm}}$ is a traceless operator, all terms proportional to $\delta_{ij}$ will vanish after contraction. Next, similar to the gauge field, we transform tensor perturbations to the Fourier space as
\begin{align}
  \notag
  \hat{h}_{ij}(\tau, \bm{x})
   =&\int \frac{\mathrm{d}^{3} \bm{k}}{(2 \MaPI)^{3 / 2}}
  \hat{h}_{ij}(\tau, \bm{k}) \\
  \notag
   =& \int \frac{\mathrm{d}^{3} \bm{k}}{(2 \MaPI)^{3 / 2}}
  \sum_{\lambda = +, -}
  (
    \sqrt{2} \epsilon_{i}^{\lambda}(\hat{k}) \epsilon_{j}^{\lambda}(\hat{k})
    \hat{a}^{\lambda}(\bm{k}) \\
    & \times h^{\lambda}(\tau, \bm{k}) \MaE^{ \MaI \bm{k} \cdot \bm{x} }
    + \text{h.c.}).
\end{align}
Introducing a projection operator $\Pi^{\pm}_{lm}(\bm{k}) \equiv \epsilon^{\mp}_{l}(\hat{k}) \epsilon^{\mp}_{m}(\hat{k}) / \sqrt{2}$, so that $h^{\pm}(\bm{k}) = \Pi_{ij}^{\pm}(\bm{k}) h_{ij}(\bm{k})$. Next, we use Green's function method to solve the EoM in the Fourier space
\begin{align}
  \notag
  \hat{h}^{\pm} (\tau, \bm{k})
  & = - \dfrac{2 H^{2}}{M^{2}_{\mathrm{pl}}}
  \int_{-\infty}^{\tau} \mathrm{d}\tau' G_{k}(\tau, \tau') \tau'^{2}
  \int \dfrac{\mathrm{d}^{3} \bm{q}}{(2 \MaPI)^{3 / 2}}
  \Pi_{lm}^{\pm}(\bm{k}) \\
  \notag
  & \times \Bigl[
    \tilde{A}'_{l}( \bm{q}, \tau' ) \hat{A}_{m}'( \bm{k} - \bm{q}, \tau') \\
  \label{eq: sourced_tensor}
    & - \varepsilon_{lab} q_{a} \hat{A}_{b}(\bm{q}, \tau')
    \varepsilon_{mcd}( k_{c} - q_{c}) \hat{A}_{d} (\bm{k} - \bm{q}, \tau')
  \Bigr],
\end{align}
where the Green's function is given by
\begin{align}
  \notag
  G_{k}(\tau, \tau') =& \dfrac{1}{k^{3} \tau'^{2}}
  \bigl[
    (1 + k^{2} \tau \tau') \sin(k (\tau - \tau')) \\
  \label{eq: gw_green_func}
    & - k (\tau - \tau') \cos(k (\tau - \tau')) \bigr] \Theta(\tau - \tau').
\end{align}
With tensor perturbations, one can compute the two-point correlation $\braket{h h}$ and the GW power spectrum $\mathcal{P}_{h}$ at a given time $\tau$ via
\begin{equation}
  \braket{h^{\pm}(k, \tau) h^{\pm}(k', \tau)}
  \equiv \dfrac{2 \MaPI^{2}}{k^{3}} \mathcal{P}^{\pm}_{h}(k, \tau) \delta(\bm{k} + \bm{k}').
\end{equation}
Applying Eq. \eqref{eq: sourced_tensor} and retaining only the enhanced mode $A^{-}(\bm{k})$, the tensor power spectrum at the end of the inflation is given by
\begin{widetext}
\begin{align}
  \notag
  \mathcal{P}_{h}^{\lambda}
  =& \frac{H^{4} k^{3}}{2 \MaPI^{4} M_{\mathrm{pl}}^{4}}
  \int_{0}^{\infty} q^{2} \mathrm{d} q
  \int_{-1}^{1} \mathrm{d} u
  \Biggr[
    \left| \epsilon^{\lambda}_{i}(\bm{k})
      \epsilon^{-}_{i}(\bm{-q}) \right|^{2}
    \left| \epsilon^{\lambda}_{j}(\bm{k})
      \epsilon^{-}_{j}(\bm{q} - \bm{k}) \right|^{2} \\
    \label{eq: sourced_tensor_spectrum}
    & \times \biggl| \int_{-\infty}^{0} \mathrm{d} \tau' \tau'^{2}
    G_{k}(\tau_{\mathrm{end}}, \tau')
    \biggl(
        A^{'-}(\tau', \bm{q}) A^{'-}(\tau', \bm{k}-\bm{q})
        + q |\bm{k} - \bm{q}|
        A^{-}(\tau', \bm{q})
        A^{-}(\tau', \bm{k} - \bm{q}) \biggr) \biggr|^{2} \Biggr],
\end{align}
\end{widetext}
where $u \equiv \cos \theta$, and the two norms of the polarization vector can be computed via the following property
\begin{equation}
  \left| \epsilon_{i}^{\lambda}(\hat{p})
    \epsilon_{i}^{\lambda'}(\hat{q}) \right|^{2}
  = \left( \frac{1 - \lambda \lambda' \hat{p} \cdot \hat{q}}{2} \right)^{2}.
\end{equation}
Next, one can compute the current energy density of the corresponding stochastic GW background as
\begin{equation}
  \Omega_{\mathrm{GW},0}h^2 = \frac{\Omega_{r, 0} h^{2}}{24}
  ( \mathcal{P}^{+}_{h} + \mathcal{P}^{-}_{h} ),
\end{equation}
where $\Omega_{\mathrm{r}, 0} \simeq 4.15 \times 10^{-5} / h^{2}$ denotes the current fraction of the radiation energy density to the total energy density of the Universe \cite{Garcia-Bellido:2023ser}.

\begin{figure}[tbp]
  \includegraphics[width=.48\textwidth]{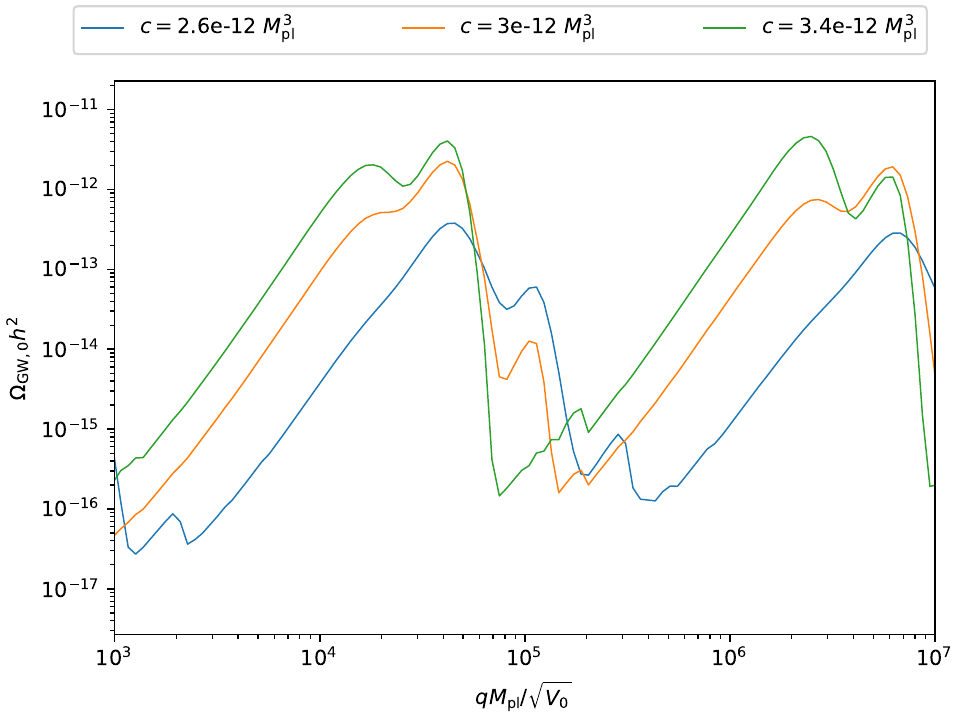}
  \caption{Energy spectrum of GWs for different slopes, where we fix the coupling constant as $f^{-1} = 400 \, M_{\mathrm{pl}}$.}
  \label{fig: OmegwGW_tot_0}
\end{figure}

Setting $\tau = \tau_{\mathrm{end}}$, for the time $\tau'$ satisfying $\tau' / \tau_{\mathrm{end}} \gg 1$, the $G\tau'^{2}$ term in the integral of Eq. \eqref{eq: sourced_tensor_spectrum} becomes
\begin{equation}
  G_{k}(\tau_{\mathrm{end}}, \tau')\tau'^{2} \simeq \frac{1}{k^{3}}
  [-\sin(k\tau') + k\tau' \cos(k\tau')].
\end{equation}
In the superhorizon region, $k / aH = -k\tau' \ll 1$, we obtain $G_{k}\tau'^{2}$ is $G_{k}\tau'^{2} \simeq -\tau^{3} / 3$. Since $\tau$ exponentially closes to zero as $t$ increases, this term will exponentially suppress the integral. On the other hand, we observe that this integral is primarily contributed by the momentum $\bm{q}$ near $\bm{k}$, i.e., $k \sim q$. If $q \ll k$, the mode $A(q)$ freezes before $A(|\bm{k}-\bm{q}|) \simeq A(k)$ starting to grow, making the $A'A'$ term too small to contribute to the integral. If $q \gg k$, when $A(q)$ begins to grow, the Green's function part is already exponentially small, strongly suppressing the integral. Therefore, only when $\bm{q}$ is in the neighborhood of $\bm{k}$, the growing parts of $A(q)$ and $A(|\bm{k}-\bm{q}|)$ overlap, and Green's function does not strongly suppress the integral. Hence, one can estimate the power spectrum of the sourced GWs via $\mathcal{P}_{h}(k) \propto \left| A^{-}(k) \right|^{4}$. Fig.~\ref{fig: OmegwGW_tot_0} shows the energy spectrum $\Omega_{\mathrm{GW}, 0}$ for different slopes of the axion potential.


\section{Conclusions}
\label{sec: conclusion}

In this paper we have investigated strong backreaction of gauge quanta to the dynamics of the axion field during inflation. We considered a toy model where the axion potential is linear, which helps avoid the complexities introduced by changes in the slope. We found that, as the slope of the potential $c$ or the coupling coefficient $f^{-1}$ is increased, the backreaction term becomes comparable to the slope $U_{,\chi}$, causing the system to enter the strong backreaction regime.
In this regime, backreaction decelerates the axion field, which leads to a decrease in the gauge quanta production. Subsequently, as the backreaction weakens, the axion field accelerates once more. This cycle repeats, resulting in oscillations in $\xi$, $\rho_{\mathrm{EM}}$, $|A|^{2}$, and other related quantities.

When backreaction is weak, the spectrum of the gauge field is exponentially dependent on both the slope $c$ and the coupling coefficient $f^{-1}$, and this exponential dependency stops in the strong backreaction regime. In the strong backreaction regime, increasing the slope $c$ continues to enhance the peak value of the gauge field spectrum, but this enhancement is no longer exponential. Besides, Fig.~\ref{fig: Ne_xi_rhoEM_br_varcslope} shows increasing the slope $c$ can also approximately linearly increase the peak value of $\rho_{\mathrm{EM}}$. In contrast, increasing the coupling coefficient $f^{-1}$ has no obvious changes in the gauge field spectrum, and the energy density Fig.~\ref{fig: Ne_xi_rhoEM_br_varalp} shows increasing $f^{-1}$ can approximately linearly reduce the peak value of $\rho_{\mathrm{EM}}$. To understand the behaviors, we compared the time evolution of the parameter $\xi$ and the energy density of the gauge field $\rho_{\mathrm{EM}}$. We found that in the strong backreaction regime, both $\xi$ and $\rho_{\mathrm{EM}}$ oscillate, with a delay between them. The peak of $\xi$ always occurs earlier than the peak of $\rho_{\mathrm{EM}}$, suggesting that changes in the particle production occur after changes in the velocity of the axion field. This result is consistent with previous studies \cite{Domcke:2020zez}. When we increase the coupling coefficient $f^{-1}$ or the slope $c$, we observe that the parameter $\xi$ increases slightly, the delay between $\xi$ and $\rho_{\mathrm{EM}}$ decreases. This indicates that, although a larger coupling coefficient $f^{-1}$ or slope $c$ makes the particle production more efficient, it also accelerates the growth of backreaction, reducing the time available for efficient particle production. Consequently, in the strong backreaction regime, the gauge field spectrum no longer depends exponentially on $f^{-1}$ and $c$.
The strong backreaction condition is described by Eq. \eqref{eq: axion_decce_condition} and its validity is dirctly supported by Fig. \ref{fig: finvEB_over_slope_varalp} and Fig. \ref{fig: finvEB_over_slope_varcslope}.
The condition suggests that increasing the slope allows for a larger particle production before the axion field decelerates, and increasing $f^{-1}$ allows a smaller particle production. As a result, increasing the slope can still increase the maximum particle production, even in the strong backreaction regime. However, compared to the weak backreaction regime, this increase is much less efficient.


\begin{acknowledgments}
This work is supported in part by the National Key Research and Development Program of China Grant No. 2020YFC2201501, in part by the National Natural Science Foundation of China under Grant No. 12305057, No. 12475067 and No. 12235019.
\end{acknowledgments}


\bibliography{reference} 

\appendix


\end{document}